\documentclass{ptptex}

\usepackage{graphicx}
\usepackage{wrapft}
\newcommand{\Slash}[1]{{\ooalign{\hfil/\hfil\crcr$#1\,$}}}

\title{
Relativistic mean-field model with density-dependent meson-nucleon couplings %
}

\author{
Kenta \textsc{Minagawa}, 
Masahiro \textsc{Kawabata} and 
Koichi \textsc{Saito}%
}

\inst{
Department of Physics, Faculty of Science and Technology, \\
Tokyo University of Science, 
Noda 278-8510, Japan
}

\abst{
Within the relativistic mean-field approach, we extend the Miyazaki model, where the NN$\sigma$ and NN$\omega$ interactions are 
modified to suppress the couplings between positive- and negative-energy states of a nucleon in matter. 
Assuming appropriate density-dependence of the meson-nucleon couplings, we study 
nuclear matter and finite nuclei.  The model can reproduce the observed properties of $^{16}$O and $^{40}$Ca well. 
We also examine if the model is {\em natural}. 
}

\begin{document}

\maketitle

Recently, the relativistic mean-field approach with density-dependent meson-nucleon couplings draws 
much attention.\cite{tokifuch} 
It is an effective model for the Dirac-Brueckner-Hartree-Fock (DBHF) theory,\cite{dbhf} 
which can reproduce the saturation property of nuclear matter using the one-boson exchange potentials 
extracted from the nucleon-nucleon scattering data.  In the DBHF calculation, the relativistic effect provides 
a strong density-dependent repulsion, which is originated from the nucleon-antinucleon pair term ($Z$ graph), 
and it is vital to obtain the nuclear saturation property. 
It should be noticed that a nuclear model based on the quark substructure of a nucleon, for example, 
the quark-meson coupling (QMC) model,\cite{qmc} the quark-mean field (QMF) model,\cite{qmf} also 
gives density-dependent 
meson-nucleon couplings through the scalar field in a nuclear medium, namely the scalar polarizability.\cite{qmc} 
Thus, it seems quite natural that the meson-nucleon couplings depend on the nuclear environment. 

About a decade ago, Miyazaki\cite{miyazaki} has proposed an interesting, relativistic mean-field model for 
nuclear matter, in which the NN$\sigma$ and NN$\omega$ vertices are modified to reduce the couplings 
between positive- and negative-energy states of the in-medium nucleon (the $+-$ couplings). Although the $+-$ couplings 
play an important role in the relativistic nuclear models including nucleon-nucleus (NA) scattering (with the relativistic 
impulse approximation (RIA)) at intermediate energies, it is known that 
the effect of the coupling to negative states is 
too strong to produce the NA scattering observables at low energies.\cite{nascatt} Tjon and Wallace have remedied this problem 
by developing a generalized RIA, in which the different $+-$ couplings from the usual RIA are introduced.\cite{nascatt} 
The vertex modification studied by Miyazaki\cite{miyazaki} may enable us to include such variation of the $+-$ couplings at the 
relativistic mean-field level. The modified vertices finally result in the density-dependent 
NN$\sigma$ and NN$\omega$ couplings, which can simultaneously reproduce the nuclear matter properties and the Dirac 
scalar and vector optical potentials given by the DBHF calculation. 

In this Letter, we generalize the Miyazaki model, and 
study not only the nuclear matter properties but also single-particle energies of finite nuclei. Lastly, we discuss 
{\em naturalness} of the model.\cite{natural}


We now modify the vertices of NN$\sigma$ and NN$\omega$ couplings using the energy projection operators, 
$\Lambda^\pm(p)=(\pm \Slash{p}+M)/2M$, where $p$ is the four-momentum of a nucleon and $M$ is the mass. 
Since the vertex, $\Gamma (= I$ or $\gamma^\mu$), is expressed by
\begin{equation}
\Gamma = \Lambda^+(p^\prime )\Gamma\Lambda^+(p) + \Lambda^-(p^\prime )\Gamma\Lambda^-(p) 
                 + \Lambda ^+(p^\prime )\Gamma\Lambda^-(p) + \Lambda^-(p^\prime )\Gamma\Lambda^+(p) , 
\label{vertex1}
\end{equation}
it may be possible to vary the strength of the $+-$ couplings, introducing two parameters, 
$0 \leq \lambda_1, \lambda_2 \leq 1$, as\cite{miyazaki}
\begin{eqnarray}
\Gamma &\to& 
\lambda_1 [ \Lambda^+(p^\prime )\Gamma\Lambda^+(p) + \Lambda^-(p^\prime )\Gamma\Lambda^-(p) ] 
+ \lambda_2 [ \Lambda^+(p^\prime )\Gamma\Lambda^-(p) + \Lambda^-(p^\prime )\Gamma\Lambda^+(p) ] , \label{vertex2} \\
&=& \frac{(\lambda_1 - \lambda_2)\Slash{p}^\prime \Gamma \Slash{p} + (\lambda_1+ \lambda_2 )\Gamma M^2}{2M^2} .
\label{vertex3}
\end{eqnarray}

In the original Miyazaki model\cite{miyazaki}, $\lambda_1$ is chosen to be unity for the scalar ($I$) 
vertex, while $\lambda_2$ is unity for the vector ($\gamma^\mu$) vertex,  
because the parameters are supposed to be constants. However, in general, the strength of the 
$+-$ couplings may depend on the nuclear environment through the Pauli blocking, $Z$ graphs etc.\cite{pr100} 
To take account of those effects in the model, we here suppose that $\lambda$ simply depends on the nuclear density, $\rho_v$:
\begin{equation}
\lambda = 1 - a \left( \frac{\rho_v}{\rho_0}\right)^{b}  , \label{dd}
\end{equation}
where $\rho_0$ is the saturation density, and each $\lambda$ has two parameters, $a$ and $b$. 
Note that, in the limit 
$\rho_v \to 0$, $\Gamma$ is identical to the original form Eq.(\ref{vertex1}). 

Using the vertex (\ref{vertex3}) and the mean-field approximation for the meson fields, the Lagrangian 
density is given by\cite{miyazaki} 
\begin{eqnarray}
{\cal L}&=& 
\bar{\psi }({\Slash{\hat p}}-M)\psi -\frac{1}{2}m^2_{\sigma } \sigma^2
+\frac{1}{2}m^2_{\omega } \omega^2 \notag \\
&+& \frac{g_{\sigma}}{2M^2} 
[ (\lambda^s_1 -\lambda^s_2 )(\bar \psi \Slash{\overleftarrow{\hat p}})(\Slash{\hat p}\psi )
      +(\lambda^s_1 +\lambda^s_2 )M^2\bar {\psi }\psi  ] \sigma \notag \\
&-& \frac{g_{\omega}}{2M^2}
[ (\lambda^v_1 -\lambda^v_2 )(\bar \psi \Slash{\overleftarrow{\hat{p}}})\gamma ^0(\Slash{\hat p}\psi )
   +(\lambda^v_1 +\lambda^v_2 )M^2\bar {\psi }\gamma ^0\psi  ] \omega , \label{lag1}
\end{eqnarray}
where $\sigma$ and $\omega$ are respectively the mean-field values of the $\sigma$ and $\omega$ mesons, 
and $\lambda^{s (v)}_i$ ($i=1, 2$) is the parameter for the scalar (vector) vertex. 
The meson mass and the NN$\sigma(\omega)$ coupling constant in vacuum are respectively denoted by 
$m_{\sigma(\omega)}$ and $g_{\sigma(\omega)}$. 

Following the prescription explained in Ref.\cite{miyazaki}\,, we can construct an {\em effective} Lagrangian 
density, in which the effect of variation of the $+-$ couplings in matter is included, 
\begin{equation}
{\cal L}_{\rm{eff}}=
\bar{\psi }(\Slash{\hat{p}}-\gamma ^0 U_v -M^{\ast})\psi -\frac{1}{2}m^2_{\sigma } \sigma^2
        +\frac{1}{2}m^2_{\omega } \omega^2 , \label{lag2}
\end{equation}
where the effective nucleon mass, $M^*$, and the Dirac scalar, $U_s$, and vector, $U_v$, potentials in 
matter are defined as 
\begin{equation}
M-M^* = g^*_{\sigma} \sigma = -U_s,  \ \ \ \ \ 
U_v = g^*_{\omega} \omega , \label{mv}
\end{equation}
with the effective coupling constants 
\begin{eqnarray}
g^*_{\sigma} &=& \frac{1}{2}[(\lambda ^s_1+\lambda ^s_2)+(\lambda ^s_1-\lambda ^s_2)(m^{*2}-v^2)]
g_{\sigma} , \label{ccs} \\
g^*_{\omega} &= & \frac{1}{2}[(\lambda ^v_1+\lambda ^v_2)-(\lambda ^v_1-\lambda ^v_2)(m^{*2}-v^2)] 
g_{\omega} , \label{cco}
\end{eqnarray}
$m^* = M^*/M$ and $v=U_v/M$. Note that, when $\lambda ^s_1=\lambda ^v_2=1$, the effective coupling constants 
coincide with those in the Miyazaki model.

The energy per nucleon, $W$, for symmetric nuclear matter is then written by
\begin{eqnarray}
W &=& \frac{3}{4}E^{\ast}_F+\frac{1}{4}M^{\ast}\frac{\rho _s}{\rho_v}+U_v-M  
+ \frac{2M}{C_s \hat \rho }\left[\frac{1-m^{\ast}}{\lambda ^s_1+\lambda ^s_2
+(\lambda ^s_1-\lambda ^s_2)({m^{\ast}}^2-v^2)}\right]^2 \notag \\
&-&\frac{2M}{C_v \hat \rho }\left[\frac{v}{\lambda ^v_1+\lambda ^v_2
-(\lambda ^v_1-\lambda ^v_2)({m^{\ast}}^2-v^2)}\right]^2 , \label{w}
\end{eqnarray}
where $E^{\ast}_F =(k_F^2+M^{\ast 2})^{1/2}$ ($k_F$ the Fermi momentum), $\rho_v = 2k^3_F/3 \pi^2$, 
${\hat \rho} = \rho_v /\rho _0$, $C_{s(v)} = g^2_{\sigma (\omega)}\rho _0/m^2_{\sigma (\omega)}M$, and 
$\rho_s = (M^*/\pi^2)[k_F E^{\ast}_F-M^{\ast 2}\ln((k_F+E^{\ast}_F)/M^{\ast})]$ (the scalar 
density). 

From the self-consistency conditions, $ (\partial W/\partial m^{\ast}) =0$ and $(\partial W/\partial v) =0$, 
which the meson fields should satisfy, one finds 
\begin{equation}
C_s =\frac{4c}{a_s^3b_s\hat{\rho}}(1-m^{\ast}), \ \ \ \ \ 
C_v =\frac{4c}{a_v^3b_v\hat{\rho}}v , \label{cc2}
\end{equation}
where
\begin{eqnarray}
a_s &=& \lambda ^s_1+\lambda ^s_2+(\lambda ^s_1-\lambda ^s_2)({m^{\ast}}^2-v^2) , \label{as}\\
a_v &=& \lambda ^v_1+\lambda ^v_2-(\lambda ^v_1-\lambda ^v_2)({m^{\ast}}^2-v^2) , \label{av}\\
b_s &=& \left[\lambda ^v_1+\lambda ^v_2-(\lambda ^v_1-\lambda ^v_2)({m^{\ast}}^2+v^2)\right]
\frac{\rho _s}{\rho_v}-2(\lambda ^v_1-\lambda ^v_2)m^{\ast}v , \label{bs}\\
b_v &=& \lambda ^s_1+\lambda ^s_2-(\lambda ^s_1-\lambda ^s_2)({m^{\ast}}^2-2m^{\ast}+v^2)
+2(\lambda ^s_1-\lambda ^s_2)(1-m^{\ast})v\frac{\rho _s}{\rho_v} , \label{bv}\\
c &=& [\lambda ^s_1+\lambda ^s_2-(\lambda ^s_1-\lambda ^s_2)({m^{\ast}}^2-2m^{\ast}+v^2)]
[\lambda ^v_1+\lambda ^v_2-(\lambda ^v_1-\lambda ^v_2)({m^{\ast}}^2+v^2)] \notag\\
     &+& 4(\lambda ^s_1-\lambda ^s_2)(\lambda ^v_1-\lambda ^v_2)(1-m^{\ast})m^{\ast}v^2  . \label{c}
\end{eqnarray}


Giving the $\lambda$ parameters, the coupling constants in vacuum, $g_{\sigma}$ and $g_{\omega}$, are determined so as to 
fulfill the saturation condition, $\partial W/\partial \hat{\rho}|_{\hat{\rho}=1}= 0$ and 
$W(\hat{\rho}=1) = -15.75\text{MeV}$ ($\rho_0 = 0.15$ fm$^{-3}$). 
Using those coupling constants, one can calculate the effective nucleon 
mass ($m^*$) and the vector potential ($v$) at any density. 

In the present calculation, we, however, set that $\lambda_{1}^s = \lambda_{2}^v = 1 - a_A ( \rho_v/\rho_0)^{b_A}$ and 
$\lambda_{2}^s = \lambda_{1}^v = 1 - c_A ( \rho_v/\rho_0)^{d_A}$ to reduce the total number of parameters.  (We call this "type A".)
Thus, for type A the density-dependence of the NN$\sigma$ coupling is  
identical to that of the NN$\omega$ coupling, i.e., $g_{\sigma}^*/g_{\sigma} = g_{\omega}^*/g_{\omega}$ 
(see Eqs.(\ref{ccs}) and (\ref{cco})).  
This choice may be justified, because it has already been found in Ref.\cite{miyazaki}\,
that the case of $\lambda_{1}^s = \lambda_{2}^v$ and 
$\lambda_{2}^s = \lambda_{1}^v$ gives the best result for the nuclear matter properties.  In contrast, we 
shall also study an alternative: $\lambda_{1}^s = \lambda_{1}^v = 1 - a_B ( \rho_v/\rho_0)^{b_B}$ and 
$\lambda_{2}^s = \lambda_{2}^v = 1 - c_B ( \rho_v/\rho_0)^{d_B}$. 
(We call this "type B".)  Thus, each type eventually has four parameters ($a \sim d$) to fit the 
observed data.   

\begin{wraptable}{l}{\halftext}
\caption{Parameter sets for type A and B. The nuclear incompressibility $K$ (in MeV) is also shown. }
\label{table:paramset}
\begin{center}
\begin{tabular}{c|cccc|c} \hline \hline
type & $a$ & $b$ & $c$ & $d$ & $K$ \\ \hline
A    & 0.15 & 0.9 & 0.26 & 0.1 & 239.3 \\
B  & 0.40 & 0.3 & 0.37 & 0.3 & 202.9 \\ \hline
\end{tabular}
\end{center}
\end{wraptable}
Now we are in a position to show our numerical results. 
We determine the coupling constants, $g_\sigma$ and $g_\omega$, so as to exactly reproduce 
the saturation condition of nuclear matter.  In addition, 
the parameters, $a \sim d$, are tuned so as to produce 
the scalar $U_s$ and vector $U_v$ potentials of the DBHF calculation and 
the observed incompressibility ($K = 210 \pm 30$ MeV) as precisely as possible. 

We find that $g_\sigma=12.24 (16.97)$ and $g_\omega=14.97 (20.69)$ for type A (B). 
In Table~\ref{table:paramset}, we list the parameter sets for type A and B. 
\begin{figure}[htb]
 \parbox{\halftext}{\centerline{\includegraphics[width=7.5 cm]{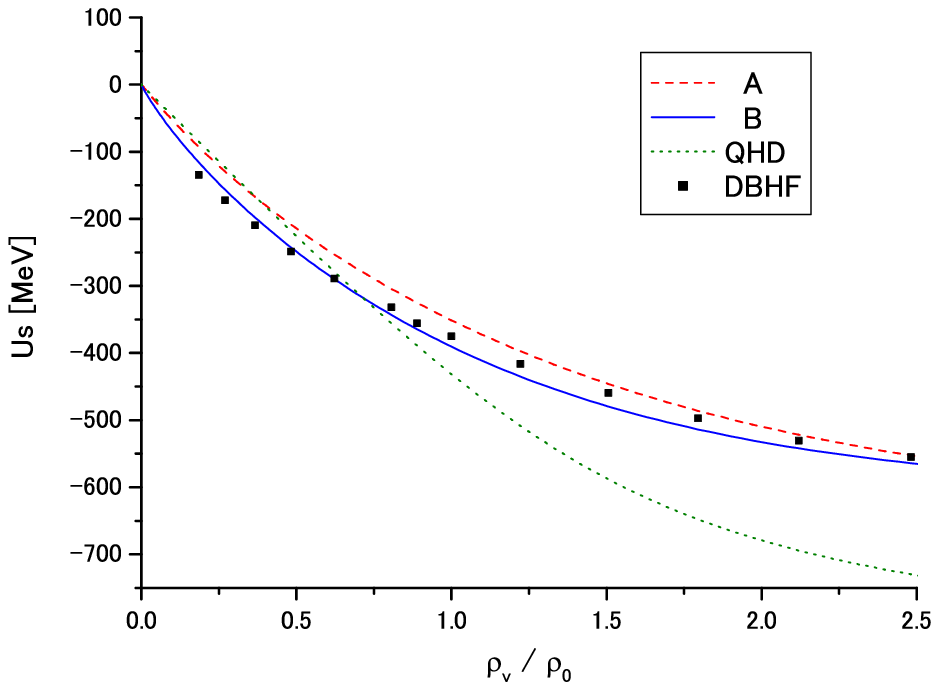}}
                \caption{The scalar potential. The dashed, solid and dotted curves are, respectively, for type A, 
B and QHD. The DBHF result is shown by solid squares. }}
 \hfill
 \parbox{\halftext}{\centerline{\includegraphics[width=7.5 cm]{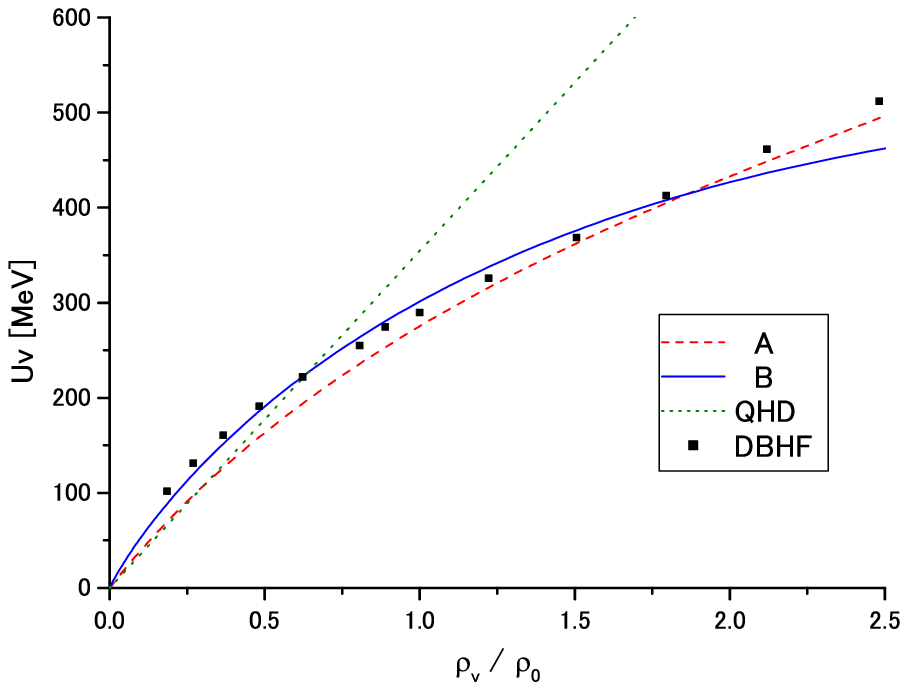}}
                \caption{The vector potential. The curves are labeled as in Fig.~1. }}
\end{figure}
In Figs.~1 and 2, the scalar and vector potentials calculated with the present parameter sets are shown, 
together with the results of DBHF calculation\cite{dbhf} and Quantum Hadrodynamics (QHD)\cite{qhd}. 
The DBHF result is well reproduced by the present model up to $\rho_v/\rho_0 = 2.0 \sim 2.5$. 
We again see from Table~\ref{table:paramset} that $g_{\sigma}^*/g_{\sigma} \simeq g_{\omega}^*/g_{\omega}$ even in type B because 
$b_B = d_B$ and $a_B \approx c_B$. Thus, for fitting the scalar and vector potentials of the DBHF calculation and 
the observed incompressibility simultaneously, 
it may be favorable that the density-dependence of the NN$\sigma$ interaction is very close to 
that of the NN$\omega$ interaction. 

\begin{table}
\begin{center}
\caption{Binding energy per nucleon $W$ (in MeV), rms charge radius $r_{ch}$ (in fm) and difference between nuclear radii for 
neutrons and protons $r_n-r_p$ (in fm). The QHD result is also included. }
\label{table:finite}
\begin{tabular}[ht]{cc|ccc|ccc}
\hline\hline
     &              &          & $^{40}$Ca &               &                & $^{16}$O &           \\
     & $m_{\sigma}$(MeV) & $r_{\text{ch}}$ & $W$       & $r_n-r_p$     &  $r_{\text{ch}}$ & $W$      & $r_n-r_p$ \\ \hline
A   & 463.4        &   3.482           & 5.78      & -0.080        &  2.83            & 4.22     & -0.047    \\
B  & 513.0        &   3.482           & 7.33      & -0.074        &  2.77            & 6.06     & -0.041    \\
QHD  & 523.8        &   3.482           & 6.24      & -0.055        &  2.75            & 4.85     & -0.033    \\
Exp. &              &   3.482           & 8.45      & 0.05$\pm$0.05 &  2.73            & 7.98     & 0         \\ \hline
\end{tabular}
\end{center}
\end{table}
\begin{table}
\begin{center}
\caption{Model predictions for the energy spectrum of $^{40}$Ca. }
\label{table:40Ca}
\begin{tabular}{c|cccc|cccc}
\hline\hline
           &       & Proton &              &       &      & Neutron &      &      \\
           &    A &   B &  QHD &      Exp. &   A &  B &  QHD & Exp. \\ \hline
1s$_{1/2}$ & 45.8 &   48.5 & 46.7 & 50$\pm$10 &  55.1 &   57.6 & 55.0 & 51.9 \\
1p$_{3/2}$ & 30.2 &    32.9 & 30.8 &  34$\pm$6 &  38.7 &    41.3 & 38.7 & 36.6 \\
1p$_{1/2}$ & 26.7 &   29.2 & 25.3 &  34$\pm$6 &  35.2 &    37.7 & 33.2 & 34.5 \\
1d$_{5/2}$ & 15.6 &  17.9 & 15.2 &      15.5 &  23.3 &     25.6 & 22.6 & 21.6 \\
2s$_{1/2}$ & 11.5 &  12.6 &  7.2 &      10.9 &  19.2 &     20.4 & 14.4 & 18.9 \\
1d$_{3/2}$ & 10.2 &  11.9 &  6.7 &       8.3 &  17.9 &     19.6 & 14.1 & 18.4 \\ \hline
\end{tabular}
\end{center}
\end{table}
\begin{wrapfigure}{l}{8.0cm}
 \centerline{\includegraphics[width=8.0 cm]{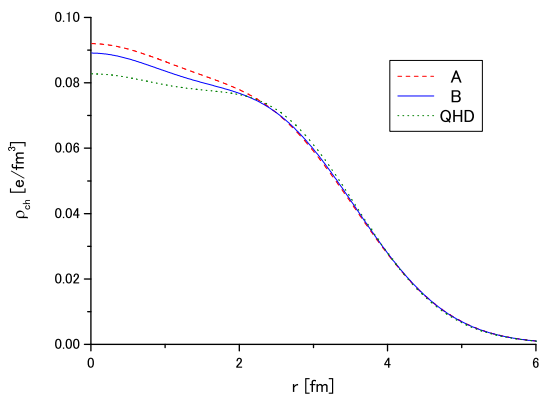}}
\caption{Charge density distribution for $^{40}$Ca compared with that of QHD. The curves are labeled as in Fig.~1. }
\label{fig:40Ca}
\end{wrapfigure}
For finite nuclei, Eq.(\ref{lag2}) gives a set of coupled non-linear differential equations, which may be 
solved by a standard iteration procedure\cite{qmc1}. For example, we have calculated the properties of $^{16}$O and $^{40}$Ca, and 
the result is presented in Table~\ref{table:finite}. 
In the calculation, we have adjusted the $\sigma$ mass ($m_\sigma$) so as to yield the observed root-mean-square (rms) 
charge radius of $^{40}$Ca: $r_{ch}(^{40}\rm{Ca}) = 3.48$ fm. 
In Table~\ref{table:40Ca}, we give the single-particle energies for $^{40}$Ca. 
We can see from the table that the model, 
especially type B, produces the good result. The spin-orbit force in the present model is thus sufficient to reproduce 
the observed energy levels and it is comparable to that of QHD. The charge density distribution for $^{40}$Ca is also 
illustrated in Fig.~\ref{fig:40Ca}, together with the QHD result. Note that the observed distribution, which is not shown here, is very 
close to that of type B.  

Finally, we shall examine the present model using Georgi's "naive dimensional analysis" (NDA).\cite{natural,nat} 
In general, an effective field theory at low energy will contain an infinite number of interaction terms, which incorporate 
the compositeness of the low-energy degrees of freedom, i.e., hadrons, and it is then expected to involve numerous couplings 
which may be nonrenormalizable. The NDA gives a systematic way to manage such complicated, effective field theories. After 
extracting the dimensional factors and some appropriate counting factors using NDA, the remaining {\em dimensionless} coefficients 
are all assumed to be of order {\em unity}. This is the so-called {\em naturalness assumption}. If theory is {\em natural}, one 
can then control the effective Lagrangian, at least at the tree level.  In the present case, 
the model involves the NN$\sigma$ and NN$\omega$ interactions.  Using NDA, we then find that the dimensionless coefficients 
corresponding to those couplings are all smaller than $2.0$.  Thus, the model is {\em natural}. 

In summary, we have extended the Miyazaki model\cite{miyazaki}, where the NN$\sigma$ and NN$\omega$ couplings are modified to suppress 
the $+-$ couplings, and studied two (A and B) types of density-dependent meson-nucleon vertices.  
Assuming an appropriate density form at the vertex, the parameters are adjusted so as to produce the scalar and vector potentials of 
the DBHF calculation and the observed incompressibility $K$ as precisely as possible.  The density-dependence such as 
$g_{\sigma}^*/g_{\sigma} \approx g_{\omega}^*/g_{\omega}$ may be eventually favorable for fitting the DBHF result and 
the observed nuclear data. Using such coupling constants, we have studied the properties of nuclear matter and finite nuclei 
($^{16}$O and $^{40}$Ca).  
The model can reproduce the experimental data well. Furthermore, NDA tells us that the model is {\em natural}. 
It is thus vital to include appropriate 
density-dependence of the meson-nucleon interactions in the relativistic mean-field approach, which may be attributed to 
many-body effects (like the Pauli exclusion) and the quark substructure of an in-medium nucleon.\cite{qmc}


%

\end{document}